\documentclass[12pt]{article}
\usepackage[utf8]{inputenc}
\usepackage{graphicx}
\usepackage{booktabs}
\usepackage{cite}

\title{Evolvability Tradeoffs in Emergent Digital Replicators}
\author{Thomas LaBar$^{1,2,3}$, Arend Hintze$^{2,4,5}$, Christoph Adami$^{1,2,3,6}$}
\begin{document}
\maketitle
\begin{center}
$^{1}$ Department of Microbiology \& Molecular Genetics\\
$^{2}$ BEACON Center for the Study of Evolution in Action\\
$^{3}$ Program in Ecology, Evolutionary Biology, and Behavior\\
$^{4}$ Department of Integrative Biology\\
$^{5}$ Department of Computer Science and Engineering\\
$^{6}$ Department of Physics and Astronomy\\
Michigan State University, East Lansing, MI 48824\\
\end{center}
\begin{abstract}
The role of historical contingency in the origin of life is one of the great unknowns in modern science. Only one example of life exists--one that proceeded from a single self-replicating organism (or a set of replicating hyper-cycles) to the vast complexity we see today in Earth's biosphere. We know that emergent life has the potential to evolve great increases in complexity, but it is unknown if evolvability is automatic given any self-replicating organism. At the same time, it is difficult to test such questions in biochemical systems. Laboratory studies with RNA replicators have had some success with exploring the capacities of simple self-replicators, but these experiments are still limited in both capabilities and scope. Here, we use the digital evolution system Avida to explore the interaction between emergent replicators (rare randomly-assembled self-replicators) and evolvability. We find that we can classify fixed-length emergent replicators in Avida into two classes based on functional analysis. One class is more evolvable in the sense of optimizing their replication abilities. However, the other class is more evolvable in the sense of acquiring evolutionary innovations. We tie this trade-off in evolvability to the structure of the respective classes' replication machinery, and speculate on the relevance of these results to biochemical replicators.
\end{abstract}

\section{Introduction}
The science surrounding the origin of life presents an obvious difficulty to scientists: our data consist of a single example. It is unknown whether the events that followed the origin of life are deterministic consequences of life or unique to this specific origin. One question of interest is whether all life, or specifically all self-replicators, are evolvable. Evolvability (the ability to evolve) has many similar, but differing definitions~\cite{wagner1996perspective,kirschner1998evolvability,wagner2013robustness}; here, we define it as the ability to increase in fitness. 
We also distinguish between two possibilities for such fitness increases: {\em optimization}---defined as the ability to improve an already-present phenotypic trait, and {\em innovation}---the ability to evolve novel phenotypic traits.

A common assumption is that life originated with self-replicating RNA molecules~\cite{gilbert1986origin,joyce2002antiquity}. Thus, most empirical studies have focused on RNA replicators, either emergent replicators~\cite{sumper1975evidence,biebricher1986template,biebricher1993sequence} or those created by experimenters~\cite{lincoln2009self,robertson2014highly}. Experiments (both computational and biochemical) have also explored the  evolvability of RNA replicators, usually involving extensive mapping of their fitness landscapes~\cite{huynen1996smoothness,jimenez2013comprehensive,athavale2014experimental,petrie2014limits}. While 
RNA is an enticing candidate for a pre-biotic molecule, the so-called 
``RNA world hypothesis" has its own problems---foremost the ``asphaltization" that tends to befall formose carbohydrates under the expected conditions of an early Earth~\cite{Deckeretal1982,Neveuetal2013}---which has led researchers to explore 
other origins of life not necessarily RNA-based~\cite{leslie2004prebiotic,robertson2012origins}, or even to move the origin 
of life from Earth to Mars~\cite{mileikowsky2000natural}. 

In recent years, more and more theoretical models concerning the origin of life have focused on exploring the abstract concepts that could possibly be involved in any potential origination, independent of a particular biochemical system~\cite{walker2013algorithmic,England2013,adami2015information}. The field of artificial life is ideally suited to study various possibilities for the origin of life as it imagines life as it could be, not just as it is. The question of the random emergence of replicators has been addressed in various digital systems before~\cite{pargellis1996spontaneous,pargellis2001digital,adami2015entropy,mathis2015emergence}, while theoretical models have explored the factors that lead to differing evolvability~\cite{ancel2000plasticity,earl2004evolvability,draghi2010mutational}. Artificial life tools have also been used to explore the potential of evolvability in different systems~\cite{taylor1999self,smith2002fitness,bedau2003evolution,egri2003evolvability,elena2008effect,frenoy2012robustness}.

Recently, Adami used information theory to calculate the likelihood of the random emergence of a self-replicator--in a sense, the progenitor to life--without regards to a specific biochemistry~\cite{adami2015information}. Adami and LaBar tested this theory with Avida by generating billions of random avidian sequences and checking for their ability to replicate themselves~\cite{adami2015entropy}. Such an investigation is akin to studying the chance emergence of self-replicating RNA molecules~\cite{biebricher1986template}.  In previous work we explored the relationship between evolvability and self-replication using these emergent replicators~\cite{labar2015does}, and found that almost all emergent replicators were evolvable, both in terms of optimization for replication and in terms of evolving beneficial phenotypic innovations. We also discovered that these replicators came in two forms: proto-replicators and self-replicators. Proto-replicators deterministically copy themselves inaccurately, but eventually evolve into self-replicators; self-replicators, on the other hand, produce an exact copy of themselves in the absence of stochastic mutation. We also noted the possibility of an optimization-innovation trade-off in some of these replicators, especially the default Avida ancestor (a hand-written self-replicator specifically designed for evolution experiments).

Here, we extend our previous study and test a fundamental question concerning life's origins: how does the genetic composition of the first replicator determine the future evolution of life? One extreme possibility is that all emergent replicators have similar genetic compositions and thus the future evolution of life will occur in a similar manner, no matter the progenitor. On the opposite end of the spectrum is the possibility that every emergent replicator is sufficiently different from every other replicator in genetic composition. In this case, the future outcome of life may be entirely dependent on which replicator emerges first. In experiments with the artificial life system Avida, we find that the interplay between the genetic composition of the first emergent replicator and future evolutionary outcomes is between these two extremes. Emergent replicators can be classified into two distinct classes based on a functional analysis of their replication machinery. These classes differ in their ability to optimize their replication ability. However, those replication classes that display high evolvability towards optimizing replication also demonstrate low evolvability towards evolving novel phenotypic traits, and vice versa. Finally, we show that this difference in evolvability is due to differences in the replication machinery between the different replicator classes


\section{Methods}
\subsection{Avida}
In order to study the interplay between emergent replicators and evolvability, we used the Avida digital evolution platform~\cite{adami1998,ofria2009avida}. In Avida, simple computer programs (``avidians") in a population compete for the memory space and CPU time needed to replicate themselves. Each avidian consists of a genome of computer instructions, where each locus in the genome can be one of 26 possible instructions. Contained with each genome are the instructions necessary for the avidian to allocate a daughter genome, copy its genome into this new daughter genome, and divide off the daughter genome. As the replication process is mechanistic (i.e., avidians execute their genome's instructions, including those to replicate, sequentially), the speed at which replication occurs is also genome dependent. Therefore, fitness is genome-dependent, as an organism's ultimate success is determined by its replication speed in these simple environments. 

As avidians directly copy and pass their genomes to their
daughters, Avida populations also have heredity. During the copying process, errors may be introduced, resulting in mutations and population variation. Therefore, Avida populations possess heredity, variation, and differential fitness: they are an {\em instantiation} (as opposed to a simulation) of Darwinian evolution~\cite{pennock2007models} This simpler instantiation of evolution by natural selection has allowed for the exploration of many topics hard to test in biological systems, see e.g.~\cite{lenski1999genome,adami2000evolution,misevic2006sexual,clune2008natural,covert2013experiments,goldsby2014evolutionary,zaman2014coevolution}.

Avida has been explained in greater detail elsewhere (see ~\cite{ofria2009avida} for a full description); here, we will cover the details relevant for this study. An avidian consists of a variety of elements: a genome of instructions, a ``read-head" that marks which instruction should be copied, a ``write-head" that denotes where the read-head-marked instruction should be copied to, and three registers to hold integer numbers (AX, BX, and CX), among other elements. 

In order to undergo reproduction, an avidian needs to perform four operations in the following order. First, it must execute an \textit{h-allocate} instruction which allocates a fixed number of \textit{nop-A} instructions to the end of its genome (here 15, as we work exclusively with length 15 genomes). These \textit{nop-A} instructions are inert by themselves, and serve as placeholders to be replaced by actual information.
Allocation of this memory prepares the daughter genome space to receive the information from the parent. Next, the read-head and write-head must be set 15 instructions apart, allowing for the instructions in the parent genome to be copied into the daughter genome. These operations are algorithmically similar to creating a DNA replication fork at the origin of replication, preparing for the assembly of a copied sequence. Following this step, the genome must find a way of looping over the \textit{h-copy} instruction in the genome to actually copy instructions from the parent into the daughter genome, similar to the action of a DNA polymerase fusing the new (daughter) nucleotides on the former parent strand. This copying can be done by either looping through the entire genome (using the circular nature of the genome to re-use a single \textit{h-copy} command as many times as necessary to copy all instructions) or else by continuously looping over a smaller set of instructions in the genome 
(called the ``copy loop"). The latter algorithm requires marking the set of instructions (the ``replication gene") so as to control the forking of execution flow. Finally, an avidian must execute an \textit{h-divide} instruction to divide the duplicated genome into two avidians, and thus successfully reproduce.

In the experimental design used here, we used two different mechanisms to produce mutations during replication. The first mechanism produces at most one {\em point mutation} at a random locus in the genome at a rate of 0.15 mutations per genome per generation upon successful division. The second mechanism is a deterministic (or ``incipient") mutation, which occurs when the instructions in an avidian cause it to copy instructions into the daughter genome in the wrong place. Often, this results in a daughter genome with one or two \textit{nop-A} instructions at the beginning if these two loci were skipped over during the copying process, remaining instead in their pristine form. In many cases, such a faulty copy algorithm results in the offspring being non-viable; however, in some cases viable offspring are produced. Sometimes, this incipient mutation mechanism results in the phenomenon of {\em proto-replicators}, defined as those replicators that deterministically (i.e., reproducibly because genetically controlled) make an offspring different from itself~\cite{labar2015does}. Replicators that instead deterministically make an identical offspring are called self-replicators. 

In the standard experimental evolution scenario, avidians are placed into a landscape that rewards a variety of phenotypic traits. These traits commonly refer to the ability to perform certain Boolean logic calculations on binary numbers that the environment provides. During an avidian's lifespan, they can input (read from) and output (write to) binary numbers from/to the environment. Whenever a number is written, the Avida world checks if a Boolean logic calculation was performed. Successful performance results in an increase in the replication speed of that individual's descendants. In the standard Avida design (also the one used here in the experiments to explore evolvability in the sense of innovation) that environment rewards the performance of nine different calculations. This environment is usually referred to as the ``Logic-9" environment and rewards calculations such as NOT and EQUALS~\cite{lenski2003evolutionary}. The more complex the calculation performed, the greater the replication speed increase. 
The performance of such logic calculations can be viewed as an algorithmic analogue of performing different metabolizing reactions using different sugar resources (the binary numbers provided by the environment). 

\subsection{Experimental Design}
To study the evolvability of emergent replicators, we first had to generate a collection of emergent replicators. We re-analyzed a list of 10$^9$ random avidian length-15 genomes generated previously~\cite{adami2015entropy}. In order to discover replicators of fixed genome size, we set the OFFSPRING SIZE RANGE parameter in the main Avida configuration file to 1.0,
which guarantees that the \textit{allocate} command allocates exactly as much space as needed by the daughter genome. For our focus on replicators of length 15, this guaranteed that exactly 15 instructions are allocated for the daughter to receive~\footnote{In previous studies~\cite{labar2015does,adami2015entropy}, OFFSPRING SIZE RANGE was set to 2.0, which allocated an extra number of loci that typically would not be filled by the replication process, and could interfere with accurate self-replication).} 
Next, we looked at the relative abundance of proto-replicators compared to true self-replicators. We re-analyzed the above set of replicators, but made one additional parameter change: we set REQUIRE EXACT COPY to 1. Any replicator that registered as having zero fitness in this analysis would deterministically copy its genome inaccurately, and would be classified as proto-replicators. Any replicators that could reproduce themselves accurately under this treatment were classified as self-replicators.

In order to test the evolvability of the these replicators,
we performed three different experimental tests. For all experiments, we used a population size of 3600. Individuals' offspring were placed in any random cell in the 60x60 grid, thus mimicking a well-mixed environment. Point mutations were applied upon division and the genomic mutation rate was 0.15 mutations per generation. Genome size was fixed at 15 instructions. The first experiment tested the replicators' evolvability in the sense of replication optimization (optimization experiments). In this experiment, each replicator was used to seed 10 populations; these populations evolved for 10$^3$ generations. Phenotypic traits were not under selection in these experiments, that is, evolution only optimized the replication
machinery. Next, we performed evolution experiments (referred to as the ``innovation" experiments) where we rewarded individuals for evolved phenotypic traits (i.e., logic tasks) besides any increase in replicatory prowess. 

We ran the innovation experiments for 10$^4$ generations (an order of magnitude longer than the optimization experiments). These experiments were designed to decrease the likelihood that our data resulted from one of the replicator classes taking longer to evolve complex traits. Finally, we repeated the innovation experiments, but used each ancestral replicator's fittest descendant from the optimization experiments to initialize the populations (trade-off experiments). These experiments tested the presence of an optimization--innovation trade-off. 

\subsection{Data Analysis}
Once we determined the set of replicators, we first tried to determine if there were similarities in the genomes of the replicators. We realized that many of the genomes had similar instruction motifs within the sequence. To test whether different replication algorithms existed in different replicators, we used Avida's TRACE function to analyze the step-by-step execution of the replicators' genomes. This allowed us to cluster the emergent replicators into two main distinct functional replicator classes. 

To examine the results from the evolution treatments, we analyzed
the most abundant genotype at the end of each experiment. The data collected included the final evolved fitness and the number of evolved phenotypic traits. All Avida analysis was performed using Avida's Analyze Mode with settings matching those under which the relevant experiments were performed. Statistical analyses were performed using the NumPy, SciPy, and Pandas Python modules~\cite{van2011numpy,scipy,mckinney2010data}. Figures were generated using the Matplotlib Python module~\cite{hunter2007matplotlib}.  

\section{Results}
\subsection{Emergence of Replicators}
Among the one billion randomly-generated genomes, we found 75 genomes that could replicate themselves when
genome size was fixed. Of these replicators, 22 were true self-replicators in the sense that they could perfectly copy their genomes even when mutations were turned off. The remaining 53 replicators were ``proto-replicators" in the sense that they could {\em not} produce a perfect copy of their genome when the mutation rate was set to zero. This deterministic miscopying is due to the specific nature of a proto-replicator's genome (see Methods). However, these replicators still produced viable offspring that would eventually lead to a self-replicator. The discovery of these proto-replicators extends the definition of proto-replicators from~\cite{labar2015does} to include fixed-length proto-replicators.

\subsection{Replication Mechanisms of Emergent Replicators}

Upon examination of these replicators, we detected the presence of distinct instruction motifs in their genomes. These motifs consist of instructions involved in genome replication and suggested that different replicators used different replication mechanisms. To explore this possibility, we performed a step-by-step analysis of each replicator's lifestyle by looking at the execution of their genome's instructions. Avidians must perform four steps to successfully reproduce: allocate a blank daughter genome, separate their read- and write-heads, copy their genome into the daughter genome through some looping process, and divide off the daughter genome (see Methods for a fuller description). We were able to cluster the replicators into two replication classes based on a difference in two traits: (1) how they separated their read- and write-heads to copy their genome and (2) how they looped through their genome in order to copy their genome. 

We named the first class of replicators ``hc" replicators because of the {\tt hc} instruction motif they all share (see Table~\ref{avidainst} for the Avida instructions and their descriptions). This class contains 62 replicators (9 self-replicators). Only 8 of these replicators have a standard copy loop (see Methods for a definition of a copy loop), which appears at the end of their genomes; these copy loops were marked by the presence of a \textit{mov-head} ({\tt g}) instruction~\footnote{Although avidian genomes are circular, they have a defined beginning instruction where genome execution starts.}. 

\begin{table*}[htbp]
   \centering
   \begin{tabular}{@{} lll @{}} 
      \toprule
      Instruction    & Description & Symbol\\
      \midrule
nop-A    & no operation (type A) & a \\
nop-B   & no operation (type B) & b \\
nop-C   & no operation (type C) & c \\
if-n-equ & Execute next instruction only-if ?BX? does not equal complement & d\\
if-less & Execute next instruction only if ?BX? is less than its complement & e\\
if-label & Execute next instruction only if template complement was just copied & f\\
mov-head & Move instruction pointer to same position as flow-head & g\\
jmp-head & Move instruction pointer by fixed amount found in register CX & h\\
get-head & Write position of instruction pointer into register CX & i\\
set-flow & Move the flow-head to the memory position specified by ?CX? & j\\
shift-r & Shift all the bits in ?BX? one to the right & k\\
shift-l & Shift all the bits in ?BX? one to the left & l\\
inc & Increment ?BX? & m\\
dec & Decrement ?BX? & n\\
push &       Copy value of ?BX? onto top of  current stack & o\\
pop & Remove number from current stack and place in ?BX? & p\\
swap-stk & Toggle the active stack & q\\
swap & Swap the contents of ?BX? with its complement & r\\
add & Calculate  sum of BX and CX; put  result in ?BX? & s\\
sub & Calculate  BX minus CX; put result in ?BX? & t\\
nand & Perform bitwise NAND on BX and CX; put  result in ?BX? & u\\
h-copy &  Copy instruction from read-head to write-head and advance both & v\\
h-alloc & Allocate memory for offspring & w\\
h-divide & Divide off an offspring located between read-head and write-head & x \\
IO &  Output value ?BX? and replace with new input & y\\
h-search & Find complement template and place flow-head after it & z\\
      \bottomrule
   \end{tabular}
   \caption{Instruction set of the avidian programming language used in this study. The notation ?BX? implies that the command operates on a register specified by the subsequent nop instruction (for example, nop-A specifies the AX register, and so forth). If no nop instruction follows, use the register BX as a default. Table from~\cite{labar2015does}. More details about this instruction set can be found in~\cite{ofria2009avida}.}
   \label{avidainst}
\end{table*}

Not having a copy loop, the remaining hc-replicators loop through their entire genome in order to replicate, that is, they use the circular nature of their genome to achieve self-replication without a dedicated copy loop. In order to separate their read- and write-heads, replicators of this class move the entry of the AX register placed by the \textit{h-alloc} ({\tt w}) execution into the CX register through the execution of one or more instructions (in all but one case one or two \textit{swap} ({\tt r}) executions). Then, they would execute \textit{jmp-head} ({\tt h}) to move the write-head 15 positions ahead to allow for genome copying (the read-head location is usually between the first instruction and the third instruction in the genome). 

We named the second type of replicators the ``fg" replicator class (given the {\tt fg} motif they share). The class consists of 12 replicators (11 self-replicators). All of these replicators have a copy loop at the beginning of their genomes; the copy loop was marked by the presence of a \textit{if-label} and \textit{mov-head} ({\tt fg}) instruction motif. In order to separate their read- and write-heads, replicators of this class do not move the sequence length (here, the number 15) from the AX register to the CX register, as hc-replicators do. Instead, they loop through their copy loop 14 or 15 times to set the location of both the read-head and write-head to 15. Next, hc-replicators send the read-head to the origin of replication (the instruction with address 0) usually via the \textit{mov-head} ({\tt g}) instruction, although two replicators used \textit{jmp-head} ({\tt h}). Then, the read- and write-heads are a sequence length number of units apart, and copying of the genome commences through the copy loop.
 
We also found one replicator that was a hybrid of these
two classes. This replicator had a copy loop at the beginning of its genome that ended with a \textit{if-label} and \textit{mov-head} ({\tt fg}) instruction motif. However, that replicator also used \textit{swap} ({\tt r}) and \textit{jmp-head} ({\tt h}) to move its write-head to 15 in order to begin genome replication, so it uses elements of both classes. This sequence turns out to be a true self-replicator, not a proto-replicator. All replicator sequences from the three classes are listed in Table S1.

\subsection{Evolvability of Emergent Replicators}
\begin{figure}
\begin{center} 
\includegraphics[height=2.67in,width=4in]{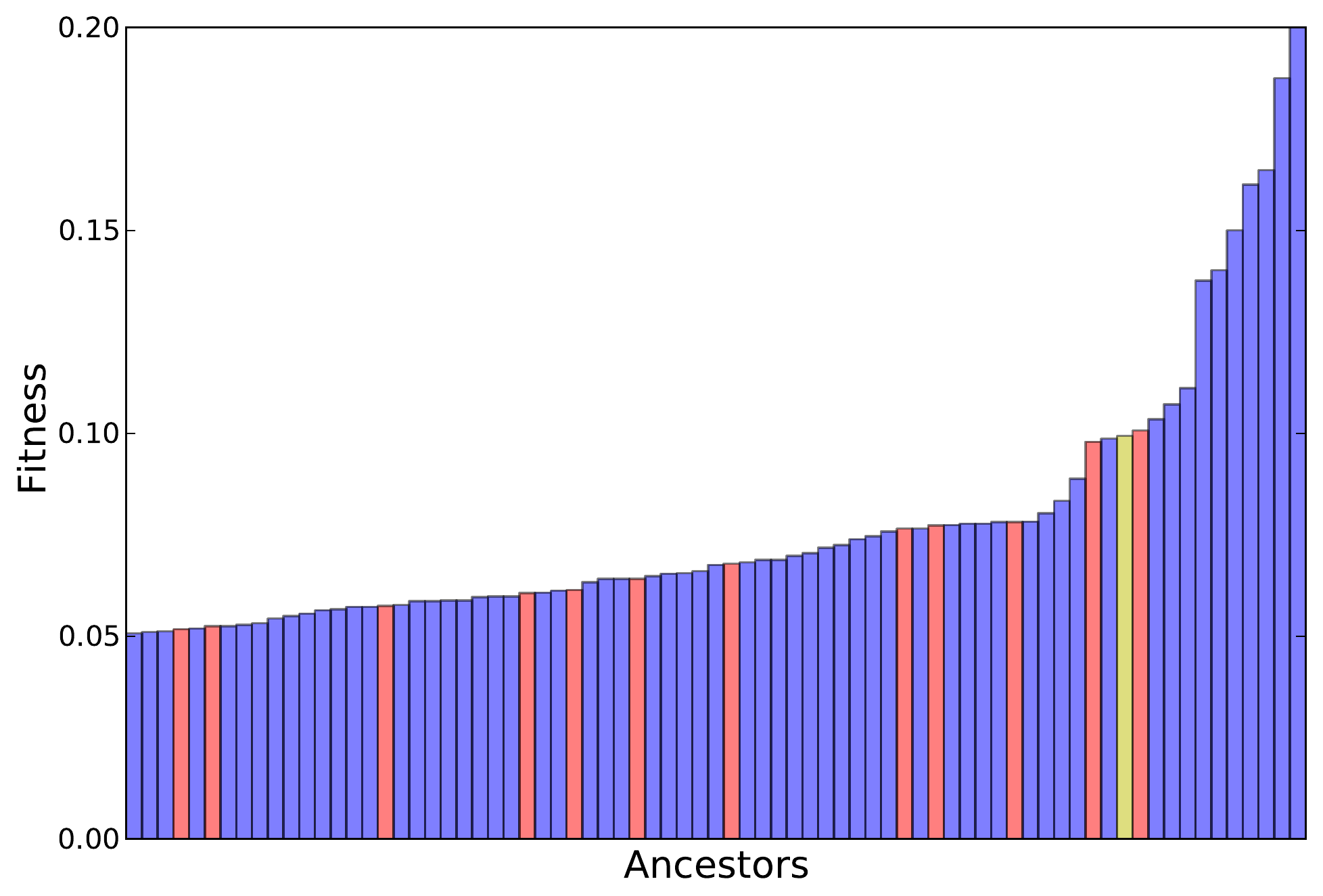}
\caption{Ancestral fitness of each emergent replicator. Blue bars are hc-replicators, while red bars are fg-replicators. The one yellow bar represents the single hybrid replicator.
}
\label{figure1}
\end{center}
\end{figure}

We performed experiments to explore the evolvability of these replicators. First, we tested how evolvable these emergent replicators are in terms of replication optimization (i.e., the optimization of the replication algorithm). The majority of replicators in any class had a fitness less than 0.1 offspring per unit of time, while, by comparison, the default Avida ancestor genotype has a fitness of approximately 0.25 offspring per unit of time (Fig.~\ref{figure1}). The replicators with the largest fitnesses were all from the hc class. 

\begin{figure}
\begin{center} 
\includegraphics[height=2.67in,width=4in]{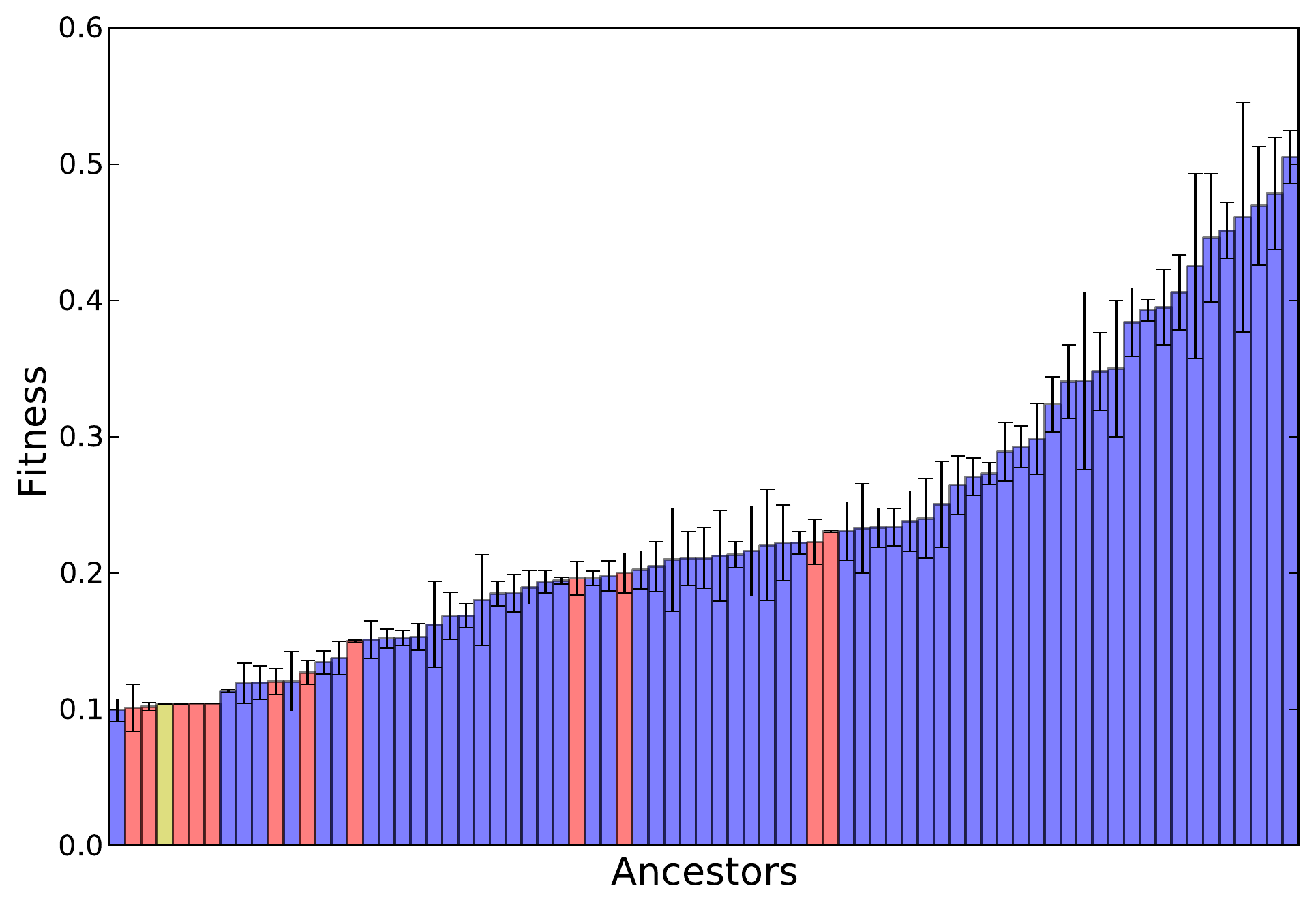}
\caption{Mean evolved fitness of emergent replicators in the optimization experiments. Error bars are twice standard error. Blue bars are hc-replicators, while red bars are fg-replicators. The one yellow bar represents the single hybrid replicator.
}
\label{figure2}
\end{center}
\end{figure}

To test for optimization evolvability, we evolved ten populations seeded with each replicator for 1,000 generations in an environment where they could not evolve new phenotypic traits (i.e., Boolean logic calculations), that is, they could only improve the speed at which they replicated. The genotypes with the highest mean evolved fitness all were of the hc type (Fig.~\ref{figure2}). Meanwhile, many of the least evolvable replicators came from the fg-replicator class. These results suggest that the replication machinery of an avidian determines how well it can optimize its replication speed, regardless of its starting fitness.

\begin{figure}
\begin{center} 
\includegraphics[height=2.67in,width=4in]{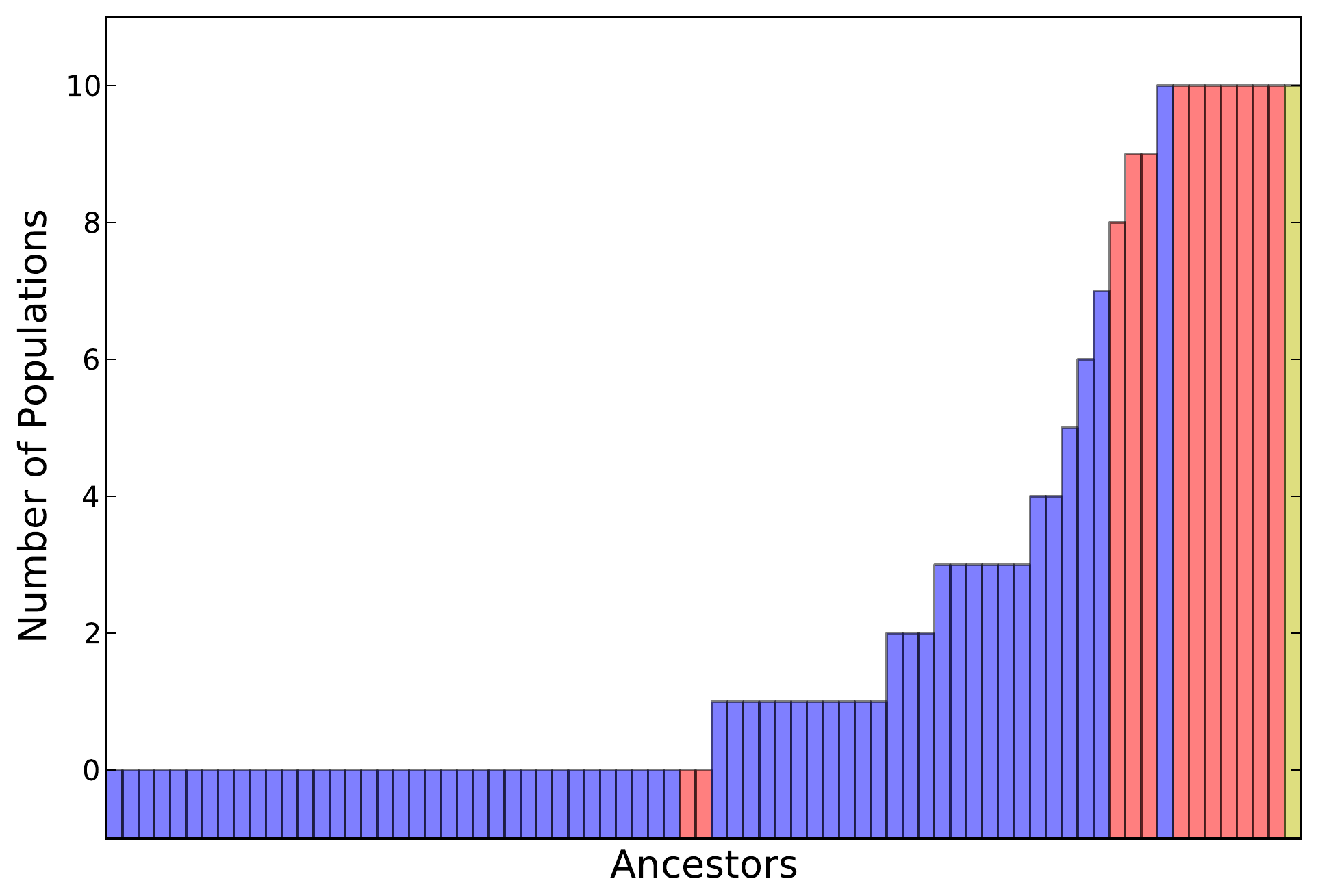}
\caption{Number of populations (out of 10) that evolved any novel phenotypic traits in the innovation experiments for each emergent replicator. Color scheme as in Fig.~\ref{figure2}.
}
\label{figure3}
\end{center}
\end{figure}

\begin{figure}
\begin{center} 
\includegraphics[height=2.67in,width=4in]{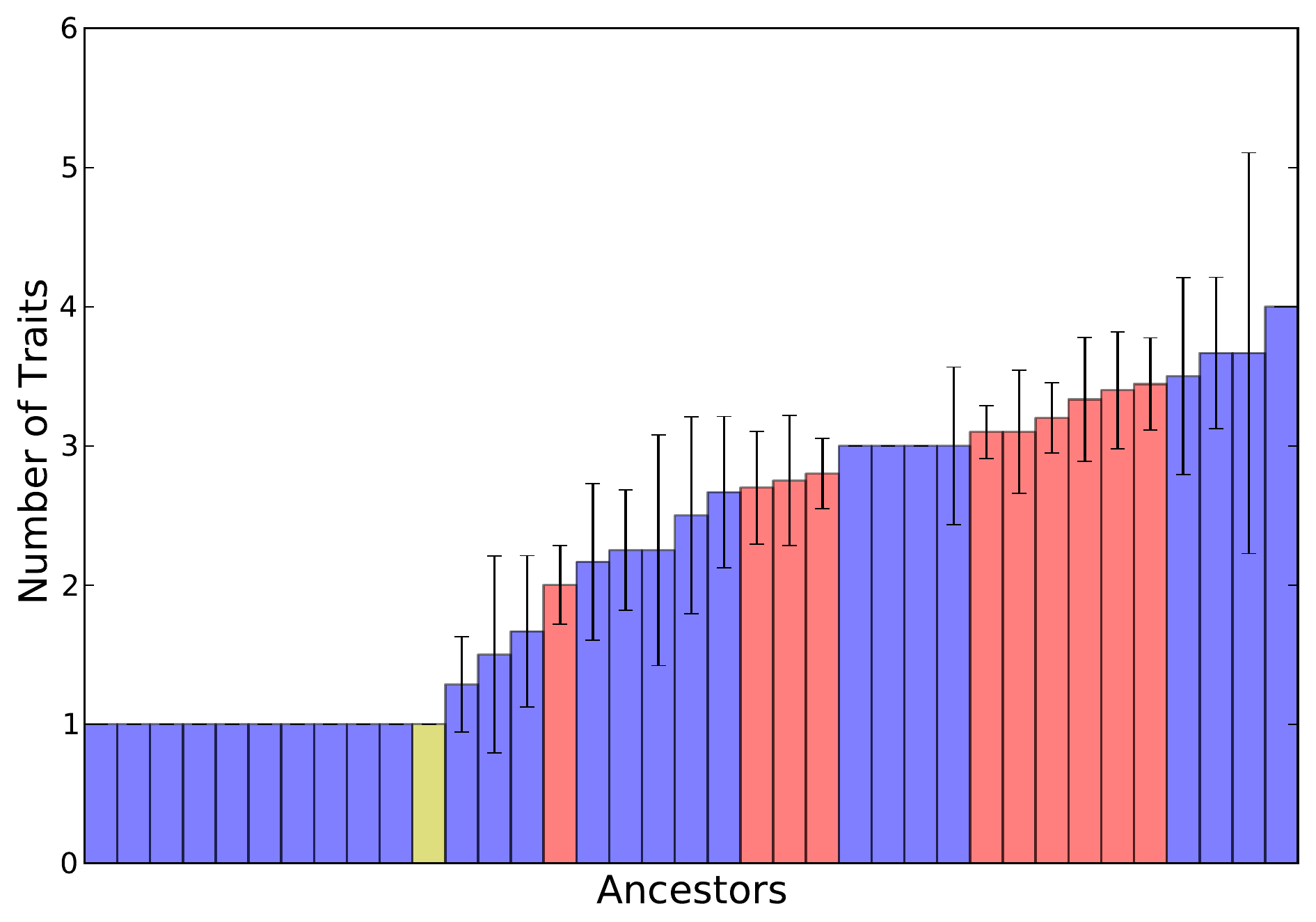}
\caption{Mean number of evolved phenotypic traits in the innovation experiments. Only populations that evolved at least one trait were included in the data. Error bars are twice standard error. Color scheme as in Fig.~\ref{figure2}.
}
\label{figure4}
\end{center}
\end{figure}

Next, we performed experiments to test how an avidian's replication class affected its ability to evolve new phenotypic traits. We again evolved ten populations of each replicator, but here we selected for the ability of these avidians to perform Boolean logic calculations (``traits"). We evolved these populations for $10^4$ generations (as opposed to the $10^3$ generations for replication optimization) to compensate for the possibility that trait evolution takes longer than optimization. Although most populations did not evolve any phenotypic traits within that time, the replicators that led to trait evolution were not evenly distributed across replication classes (Fig.~\ref{figure3}). We found that 7/12 of the fg-replicators evolved at least one trait in every replicate; only 2/12 fg-replicators did not evolve traits in any experiment. In comparison, only one hc-replicator evolved traits in every replicate, and 36 hc-replicators never evolved any traits. However, the trend was less distinct when considering the number of traits evolved in replicates that evolved at least one trait (Fig.~\ref{figure4}). We found that certain hc-replicators could evolve as many traits as fg-replicators.

\subsection{Mechanism of Evolvability Trade-off}
The above results indicate that one class of replicators is better at optimizing replication ability (the hc-replication class) and the other is better at evolving innovations (the fg-replication class). This observation suggests a trade-off between evolvability in the sense of phenotypic optimization, and evolvability in the sense of phenotypic innovation. To explore this further, we studied how the different replication mechanisms affect and determine evolvability. Specifically, we examined the respective algorithms  the two classes use, and estimated the minimal number of instructions needed to replicate (as genotypes that execute fewer instructions have a greater fitness). 

First, we explored the replication algorithm used by the fittest hc-replicators. These replicators execute their entire genome (15 instructions). During this execution step, it will allocate its daughter genome, set the write-head to 15 (the read-head is set at 0 by default), and copy $V$ instructions into the newly-allocated daughter genome, where $V$ is the number of \textit{h-copy} ({\tt v}) instructions in the genome. So far, 15 instructions have been executed and $V$ instructions have been copied. The avidian needs to copy a further $15-V$ instructions and divide off the daughter genome to reproduce. During this first genome execution, the last instruction executed is a \textit{mov-head} ({\tt g}), which moves the instruction pointer back to the beginning of that avidan's copy loop. The avidian will then loop through the copy loop until its entire genome is copied into its daughter genome. As it needs to copy a further $15-V$ instructions (and it can copy $V$ instructions per traversal) it will loop through $\frac{15-V}{V}$ times. During each traversal through the copy loop, $L$ instructions are executed, where $L$ is the size of the copy loop. Thus, to replicate the rest of the genome, $\frac{(15-V)L}{V}$ instructions are executed. Therefore, hc-replicators execute $15 + \frac{(15-V)L}{V} = 15 - L + 15\frac{L}{V}$ instructions to reproduce.

Next, we will calculate the minimal number of instruction executions needed to replicate an fg-replicator. Unlike the hc type with a copy loop at the end of their genomes, fg-replicators have a copy loop at the beginning of their genomes. Denote this copy loop size again by $L$ and the number of \textit{h-copy} ({\tt v}) instructions by $V$. First, fg-replicators will loop through their copy loop 15 times, executing $15L$ instructions during this step. During this looping, the fittest replicators will have allocated their daughter genomes. This repeated looping process sets the location of both the read- and the write-heads to 15. After the last copy execution is complete, the rest of the genome is executed, requiring $15-L$ instruction executions. During this step, the read-head location is set to 0. Next, the instruction pointer is moved back to the beginning of the genome to begin genome copying. As there are $V$ \textit{h-copy} ({\tt v}) instructions in the copy loop, the copy loop will need to be traversed $\frac{15}{V}$ times. Because the copy loop is $L$ instructions long, this step requires $\frac{15L}{V}$ instruction executions. Therefore, in total, fg-replicators require $15L + 15 - L + 15\frac{L}{V}$ instruction executions to reproduce.

By comparing the the number of instruction executions required for replication by the two classes, we can see that fg-replicators require $15L$ more instruction executions when the copy loop size and the number of \textit{h-copy} instructions are fixed. We note here that this does not mean that fg-replicators have a greater mutation rate than hc-replicators, as mutations are applied during division, not genome copying. One other factor contributes to the optimization advantage of hc-replicators. The copy loop of fg-replicators is located at the beginning of their genomes. These copy loops cannot easily expand, as they are bounded in the genome by the {\tt fg} instruction motif; expansion requires at least two mutations. However, hc-replicators have a copy loop at the end of their genome; this copy loop can more easily expand. While a copy loop expansion with a random instruction would increase the number of executions required for replication, a copy loop expansion with an additional \textit{h-copy} instruction would decrease the number of times the copy loop needs to be traversed. This would decrease the number of instruction executions needed for reproduction and increase the fitness of an hc-replicator.

Finally, we explored what prevented hc-replicators from evolving any novel traits in many populations, for example whether there is an optimization--innovation trade-off. Replicators from the hc class tend to fix beneficial mutations that enhance their replication ability. These mutations accumulate in regions of the genome that subsequently cannot be mutated into the type of instructions that discover novel traits. In turn, fg-replicators do not optimize their replication ability as well as hc-replicators, and therefore they do not (in a sense prematurely) fix as many beneficial mutations to optimize the replicator machinery. The failure to do so appears to have opened the opportunity for those sequences to instead neutrally accumulate mutations that would later lead to trait evolution. 

\begin{figure}
\begin{center} 
\includegraphics[height=2.67in,width=4in]{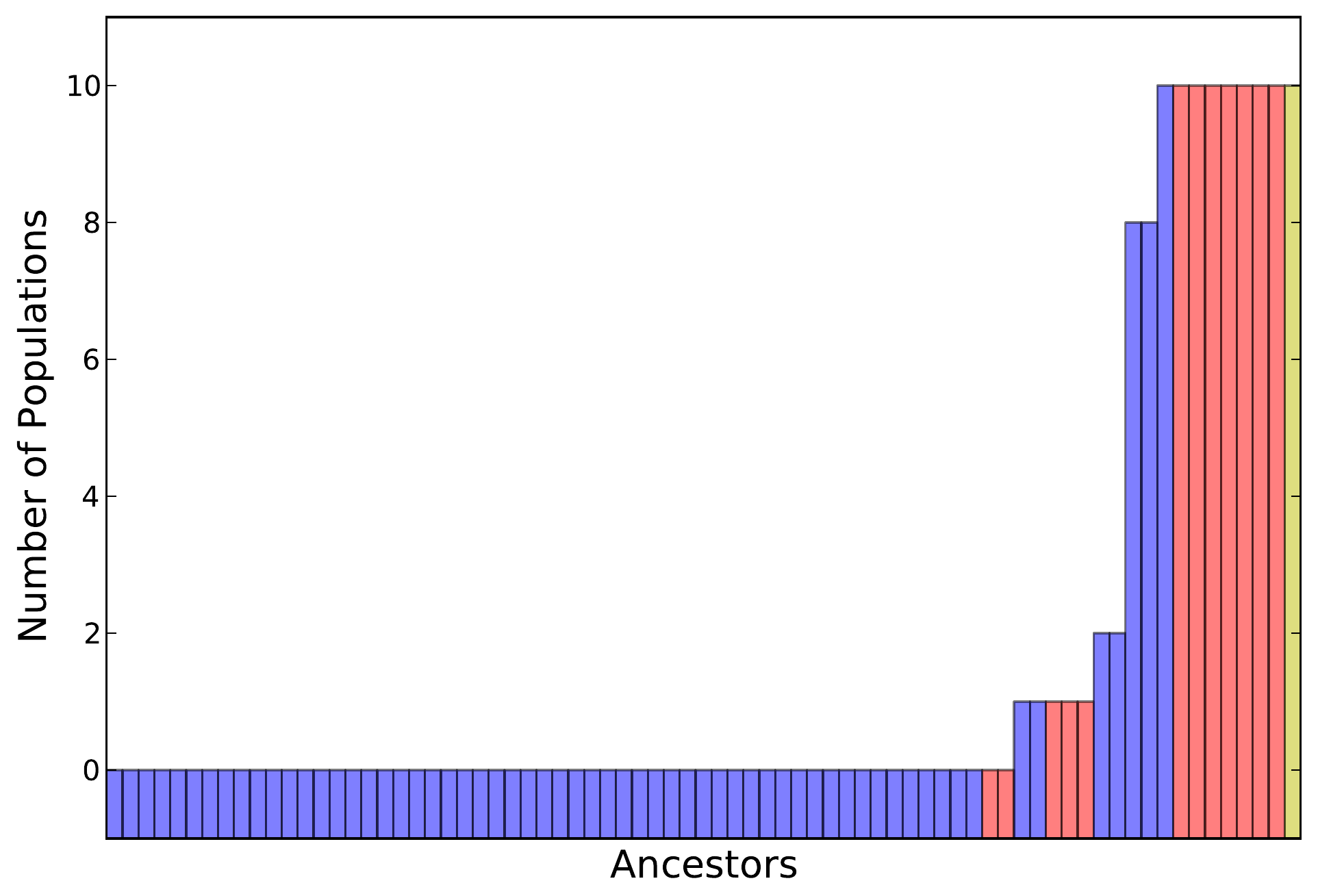}
\caption{Number of populations (out of 10) that evolved any novel phenotypic traits in the trade-off experiments for each ancestral replicator. Color scheme as in Fig.~\ref{figure2}.}
\label{figure5}
\end{center}
\end{figure}

To test this idea, we took the fittest descendant of each ancestral replicator from the optimization experiments and evolved them in an environment where phenotypic traits were under selection for $10^4$ generations (Fig.~\ref{figure5}). These experiments are the same as the innovation experiments, except we used evolved ancestors instead of the emergent replicators. Ten fg-replicators and the hybrid replicator evolved novel traits in at least one population, and many of these replicators evolved traits in all ten populations. There is little difference in the evolution of traits in the fg-replicators between these experiments and the original innovation experiments (Fig.~\ref{figure3}). However, the rate of trait evolution decreased in the hc-replicators between these experiments and the original innovation experiments. For the hc-replicators, 26/62 evolved traits in at least one population in the original innovation experiments; only 7/62 replicators evolved traits in these new innovation-after-optimization experiments. The decrease of trait evolution between the original innovation experiments and these experiments in the hc-replicators (but not the fg-replicators) indicates that their ability to optimize their replication prevents the hc-replicators from later evolving novel phenotypic traits.

\section{Discussion}
We explored the evolvability of a set of emergent (i.e., found through random search), fixed length, self-replicating sequences in the Avida digital life world. We found two distinct classes of replicators that differ in their general (algorithmic) strategy for self-replication. We found that one of these classes displayed an enhanced evolvability in terms of replication optimization, while the other was more evolvable in terms of phenotypic innovation. The experiments performed here complement our previous work on the interplay between evolvability and self-replication, where we found that most {\em variable length} replicators were evolvable both in optimization and innovation~\cite{labar2015does}. Using the present (distinct) set of replicators, we were able to provide a functional explanation for the differences of evolvability in emergent Avida replicators.

What consequences do these results have for the origin of life and subsequent evolution of said life? There is some evidence to suggest that an optimization--innovation trade-off may have existed in the earliest RNA-based replicators. In experiments with RNA bacteriophage Q$\beta$, Mills, Peterson, and Spiegelman
showed that this virus would shrink to a minimal genome and remove the genes essential for a viral lifestyle in order to optimize replication speed~\cite{mills1967extracellular}. This suggests---although certainly does not prove---that a selection pressure to minimize genome size may have been part of the RNA self-replication fitness landscape. 
When genome space is removed by selection, the ability for phenotypic innovation to occur is impeded as the remaining genomic space may become too small to contain the genes encoding novel traits.

We believe that the {\em potential} for an optimization--innovation tradeoff that hinders the evolution of phenotypic novelty may be universal. However, we also know that the earliest replicators on Earth, whether RNA-based replicators or other biochemical organisms, could eventually evolve phenotypic innovations. Furthermore, Voytek and Joyce showed that two RNA enzymes could adapt to distinct ecological niches in continuous evolution experiments, supporting the idea that novel phenotypic traits could evolve in RNA replicators~\cite{voytek2009niche}. In fact, a similar trade-off was also observed in experiments with Avidian replicators (using a single hand-written ancestor)~\cite{WhiteAdami2004}, which showed that evolution could take two different paths: one that favored fast replication, and another that traded replication speed for the exploitation of complex phenotypes. 

The optimization--innovation tradeoff we have argued for here can also be interpreted as a tradeoff for evolvability in two separate fitness components. In a previous paper, this tradeoff between replication rate and phenotypic complexity was compared to the well-known tradeoff between $r$ and $K$ reproduction strategies~\cite{WhiteAdami2004}. However, the tradeoff between fitness components in these replicators is not a fitness tradeoff, as classically discussed~\cite{stearns1989trade,roff2007evolution,hereford2009quantitative,shoval2012evolutionary}. Instead, the tradeoff is one in terms of evolvability between two fitness components. Each replicator class is biased in terms of the generation of genetic variation and this bias determines which fitness component is improved by each replicator class. 
 This result complements the previous observation in Avida, which argued that historical contingency influenced which fitness component a population improved~\cite{WhiteAdami2004}. The present results identify another factor, namely that genetic background alone may determine a population's evolutionary trajectory towards a specific ecological strategy.

One factor thought to influence evolvability is mutational robustness and the presence of neutral mutations~\cite{draghi2010mutational,masel2010robustness,wagner2013robustness}. The presence of neutral mutations results in neutral networks throughout the fitness landscape, allowing populations access to novel phenotypes~\cite{wagner2011origins}. We found a similar result here. Replicators from the fg class fixed fewer beneficial mutations that optimize replication ability. As a consequence, more of their genome's instructions were neutral, and this in turn allowed them to accumulate many mutations that eventually led to the evolution of complex traits. 

What role did the presence or absence of neutral networks play in the origin of life? Early work on (computational) RNA secondary structure predicted the existence of extensive neutral networks in RNA landscapes~\cite{huynen1996smoothness}. However, more recent experimental work suggests instead that the empirical fitness landscape of short RNAs contains isolated fitness peaks~\cite{jimenez2013comprehensive,petrie2014limits}. It is possible that different emergent replicators could either be stuck on an isolated fitness peak (such as the hc-replicators) or else have access to a neutral network that leads to phenotypic innovation (such as the fg-replicators). Such a fitness landscape architecture would emphasize a strong role for historical contingency in the origin and subsequent evolution of life.

Finally, in the work described here, avidians obtain the ``energy" necessary to replicate by performing logic calculations on random binary numbers (the ``tasks"-metabolism), but the randomly generated replicators do not have any such metabolic instructions. This is possible because an early design-decision gave each avidian a certain amount of ``base" energy for free, which they can augment by performing logic calculations. We can ask whether requiring that replicators possess some basic metabolic process (such as, for example NOT, the simplest of the logic operations) would change the conclusions we arrived at here.

Requiring a minimal metabolism would certainly increase the minimal amount of information a replicator must have from the $I=-\log_{26}{75/10^9}\approx 5.23$ instructions~\cite{adami2015information} 
by a few instructions, making the emergence of a replicator by chance that much more rare. We also would expect that the metabolic instructions would be integrated with the replicator genes, which could affect the ability of a replicator to optimize its replication machinery without affecting the metabolism. On the other hand, having a seed metabolism present could also enhance the evolution of other logic operations, for example by code duplication. While a detailed investigation of these questions must await further work, an origin-of-life scenario without metabolic genes is not altogether far-fetched. For example, the Lost City hydrothermal vents located near the summit of the Atlantis Massif~\cite{brazelton2006methane,brazelton2011metagenomic} on the mid-Atlantic seafloor produces hydrogen and simple carbohydrates ``for-free", which could be utilized by the simplest primordial replicators without the need for explicit genes. Such replicators could then evolve the genes for autonomous metabolism later on, just as the avidians do in the experiments described here. 

\section{Acknowledgements}
Computational work in support of this research was performed at Michigan State University's High Performance Computing Facility. This material is based in part upon work supported by the National Science Foundation under Cooperative Agreement No. DBI-0939454. Any opinions, findings, and conclusions or recommendations expressed in this material are those of the author(s) and do not necessarily reflect the views of the National Science Foundation.


\newpage

\onecolumn
\section*{Supplemental Material: Sequences of replicators}
\vskip 2cm
\noindent Table 1: Sequences of fixed length replicators organized by replication class
\vskip 0.5cm
\centering
   \begin{tabular}{ llll } 
      \toprule
{\bf ``hc" Genomes}\\
{\tt nuvhcwzscrchrxw}&
{\tt wdphcrxswvprcff}&
{\tt cxrphcwravgmspm}&
{\tt dwtwrvhcbxivzrr}\\
{\tt wmjrcshcikcivxg}&
{\tt ahcqdqvwzjrcxet}&
{\tt smvqztrchcwxxwj}&
{\tt vvtxwfiwvzrchch}\\
{\tt exwakshcvjjrckp}&
{\tt zezvkrcwfsshclx}&
{\tt dzzkzpvrcbwoxhc}&
{\tt kpcchcpavbcrxwo}\\
{\tt zwrchceekovxyzi}&
{\tt rwvwtzubrcxkthc}&
{\tt mzvjwmwrcuxhcng}&
{\tt jvwpwhulxdarchc}\\
{\tt pwzrvxqaarcchcl}&
{\tt upathcvuwwwxkrc}&
{\tt lxvwowxhcrcqnml}&
{\tt lbrcwavhcwaahwx}\\
{\tt yhcxxclwveblzrc}&
{\tt whcwvemcxrchoxp}&
{\tt qchcowvxrptjlbr}&
{\tt nxzjvwzxsrchcjl}\\
{\tt axzrchctvwijsca}&
{\tt swvrhcxblrpwdir}&
{\tt lvlkwovtrcxphcp}&
{\tt xshcvxwrcdetbck}\\
{\tt ynprcvxhcdfwlcj}&
{\tt ezvxwpwrchcdkqc}&
{\tt bwhcpvqzrcfhdtx}&
{\tt awexwcbvrvbxhcr}\\
{\tt qsozvdretrcwxhc}&
{\tt xlczararhchvwxq}&
{\tt izvrcqhczxbawdw}&
{\tt oovwxnhrcqhclmy}\\
{\tt ojrchcwxvfcwgzi}&
{\tt fjyrrcbwhcixvbg}&
{\tt kpavwrchcnbxupb}&
{\tt wrchcsndlzfrvxg}\\
{\tt udptwrcxvnwvhca}&
{\tt ozpvoabkwxcrchc}&
{\tt unhcwklcvzrcxjg}&
{\tt wwrcvohckqcxhvp}\\
{\tt xhwfkmrcdjhcixv}&
{\tt yzpftzvrctdhcwx}&
{\tt frmvwzkrcvxvjhc}&
{\tt slcrvhcwsslxzra}\\
{\tt aarchcekvwxtpeg}&
{\tt wifhorvthcxppwr}&
{\tt awmmqlvxhmmrchc}&
{\tt lbvxztawrcvlahc}\\
{\tt szhcwrcwvxhcfwa}&
{\tt lvxhtwjmwwrchcr}&
{\tt mphcwvadiqslxrc}&
{\tt lkexjvozxrchcwc}\\
{\tt cpvzrcqxjhcwtdj}&
{\tt bfqacswrcvxhczt}&
{\tt peenvzwrchcxjgw}&
{\tt pshcwrcpvfxqxld}\\
{\tt eazhwxrchcvodsh}&
{\tt iklkvwozljrchcx}\\

{\bf fg-Genomes}\\
{\tt vfgxqhmwmfjphgb}&
{\tt wvshxfgkfooxugb}&
{\tt kvxfgwfyxukujgb}&
{\tt dywqvphfguxqdgb}\\
{\tt vmfifgwvpowxgbt}&
{\tt aytvvmwxkfgohbi}&
{\tt xnkwovyfgtxwqgb}&
{\tt vnfgudsftwxwhgb}\\
{\tt vlfgvmhwuxwlgbq}&
{\tt cxvwfgepkhbtshi}&
{\tt wvfghqtzoxjirgb}&
{\tt wvfgxdmoprllwgb}\\
{\bf Hybrid Genome}\\
{\tt vmfgqwrmdkyxuhc}\\
      \bottomrule
      
    \end{tabular}
  
   \label{tab:seq}

\end{document}